\documentclass[preprint,12pt]{revtex4}
\usepackage{amsfonts}
\usepackage{amsmath}
\usepackage{amssymb}
\usepackage{graphicx}

\begin{document}

\parskip 12pt
 \parindent 12pt
 \baselineskip 18pt
 \hyphenpenalty 9900 \exhyphenpenalty 9900 \overfullrule 0pt

\title{Decoherence induced by magnetic impurities in quantum Hall system}
\author{V. Kagalovsky$^{1,3}$ and A. L. Chudnovskiy$^{2,3}$} 
\affiliation{$^1$ Shamoon College of Engineering, 
Bialik/Basel St., Beer-Sheva 84100, Israel \\
$^2$I. Institut f\"ur Theoretische Physik, Universit\"at Hamburg,
Jungiusstr 9, D-20355 Hamburg, Germany \\ 
$^3$Max-Planck-Institut f\"ur Physik komplexer Systeme
N\"othnitzer Str. 38,  01187 Dresden, Germany}

\begin{abstract}
Scattering by magnetic impurities is known to destroy coherence of electron motion in metals and semiconductors. 
We investigate the decoherence introduced in a single act of electron scattering by a magnetic impurity in a quantum Hall system. To this end we introduce a fictitious nonunitary  scattering matrix $\mathcal{S}$ for electrons that reproduces the exactly calculated scattering probabilities. The strength of decoherence is identified by the deviation of eigenvalues of the product $\mathcal{S}\mathcal{S}^{\dagger}$ from unity.  Using the fictitious scattering matrix, we estimate the width of the metallic region at the quantum Hall effect inter-plateau transition and its dependence on the exchange coupling strength and the degree of polarization of magnetic impurities. 
\end{abstract}


\maketitle

\section{Introduction}

Scattering by magnetic impurities can affect transport properties of electron systems substantially. Apart from the prominent Kondo effect, magnetic impurities provide a strong source of decoherence at temperatures exceeding the Kondo temperature \cite{Glazman-Kaminski,Kumar2004}. The decoherence effect is manifested especially strongly in the suppression of Anderson localization in disordered systems \cite{Ramakrishnan,Altshuler-Aronov}. In particular, scattering by magnetic impurities can create a finite metallic region near the inter-plateaux transition in the integer quantum Hall effect (IQHE) \cite{kag-ch11}. The characterization of the degree of decoherence introduced by magnetic impurities and evaluation of the corresponding phase coherence length provide an important information for interpretation of transport experiments. 
In presence of decoherence, the dynamics of the physical system ceases to be unitary \cite{LidarWhaley2003}.  
In this paper we introduce a measure of decoherence based on the nonunitarity of a fictitious scattering matrix constructed after the averaging the scattering probabilities over magnetic impurities. 

Our paper is organized as follows. In Section \ref{toyModel} we consider a toy model and show that the nonunitarity of the scattering matrix is related to the uncertainty of the phase of the wave function.   The exact scattering matrix for an electron in the saddle point potential in the quantum Hall regime and in presence of magnetic impurities is calculated in Section \ref{exactSolution}. 
Our main results are given in Sections \ref{S_fictitious} and \ref{plateaux}, where we calculate the fictitious scattering matrix, use it to determine the degree of decoherence induced by magnetic impurities, and finally estimate the width of the inter-plateaux transition. 
In Section \ref{sum} we summarize our results and discuss possible further applications of the presented method.

\section{Nonunitarity of scattering matrix as a measure of decoherence} 
\label{toyModel}

In this section we show on a simple illustrative example that the deviation of eigenvalues of the product $\mathcal{S}\mathcal{S}^{\dagger}$ (where $\mathcal{S}$ is the fictitious scattering matrix)  from unity serves as a measure of a decoherence introduced by scattering. 
To this end, we consider a simple scattering problem with two-dimensional Hilbert space. The two orthogonal incoming states are parametrized as  
\begin{equation}
 \psi_1=\frac{1}{\sqrt{\pi}}\cos\varphi, \  \psi_2=\frac{1}{\sqrt{\pi}}\sin\varphi, 
\label{in-states}
\end{equation}
and the scalar product is defined as an integral over the angle $\varphi$, 
\begin{equation}
\langle\psi_i | \psi_j\rangle= \int_0^{2\pi}\psi_i^*(\varphi)\psi_j(\varphi)d \varphi. 
\label{ScalarProduct}
\end{equation}
We assume that in the act of scattering the states experience both a potential scattering that is described by the transmission amplitude $t$ and the reflection amplitude $r$, ($r^2+t^2=1$), and random phase shifts $\alpha_1$ and $\alpha_2$ that describe the decoherence effect. Thus the outgoing states are given by 
\begin{eqnarray}
& & 
\tilde{\psi}_1^{\mathrm{out}}=\frac{1}{\sqrt{\pi}}\left\{r\cos(\varphi+\alpha_1) +  t\sin(\varphi+\alpha_2)\right\}, 
\label{psi1_out} \\ 
&& 
\tilde{\psi}_2^{\mathrm{out}}=\frac{1}{\sqrt{\pi}}\left\{-t\cos(\varphi+\alpha_1) +  r\sin(\varphi+\alpha_2)\right\}. 
\label{psi2_out}
\end{eqnarray}
In this model, the decoherence violates the orthogonality of the outgoing states.  The completely coherent scattering is realized in the case $\alpha_1=\alpha_2$. The degree of decoherence grows with the difference $\alpha_1-\alpha_2$. It is maximal for $\alpha_1-\alpha_2=\pm \pi/2$, when initially orthogonal states become linearly dependent after the scattering.  In the notations employed, a state goes in itself by coherent reflection (amplitude $r$), and it goes into the other state by coherent transmission (amplitude $t$). Now let us introduce a (nonunitary) scattering matrix for an incoherent scattering process according to the relation 
\begin{equation}
\left(\begin{array}{c}
       \tilde{\psi}_1^{\mathrm{out}} \\
	\tilde{\psi}_2^{\mathrm{out}}
      \end{array}
\right) = \mathcal{S}_{\mathrm{incoh}}\left(\begin{array}{c}
       \psi_1 \\
	\psi_2
      \end{array}
\right)= \left(
\begin{array}{cc}
 \tilde{r}_1 & \tilde{t}_1 \\
-\tilde{t}_2 & \tilde{r}_2 
\end{array}
\right)
\left(\begin{array}{c}
       \psi_1 \\
	\psi_2
      \end{array}
\right) 
\label{fict_S-example}
\end{equation}
Comparison with Eqs. (\ref{psi1_out}), (\ref{psi2_out}) allows to identify the elements of the matrix $\mathcal{S}_{\mathrm{incoh}}$ as 
\begin{eqnarray}
&& 
\tilde{r}_1=\langle \tilde{\psi}_1^{\mathrm{out}}|\psi_1\rangle=r\cos\alpha_1+t \sin\alpha_2, \\ 
&& 
\tilde{t}_1=\langle \tilde{\psi}_1^{\mathrm{out}}|\psi_2\rangle=t\cos\alpha_2 - r\sin\alpha_1, \\
&& 
\tilde{t}_2=\langle \tilde{\psi}_2^{\mathrm{out}}|\psi_1\rangle=t\cos\alpha_1 -r \sin\alpha_2, \\
&& 
\tilde{r}_2=\langle \tilde{\psi}_2^{\mathrm{out}}|\psi_2\rangle=r\cos\alpha_2+t \sin\alpha_1. 
\end{eqnarray}
The deviation of the scattering matrix $\mathcal{S}_{\mathrm{incoh}}$ from unitarity can be characterized by the products of this matrix with its hermitian conjugated. We note that for incoherent scattering the matrices $\mathcal{S}_{\mathrm{incoh}}$ and $\mathcal{S}_{\mathrm{incoh}}^{\dagger}$ do not commute any more. However, explicit calculation shows, that the products $\mathcal{S}_{\mathrm{incoh}} \mathcal{S}_{\mathrm{incoh}}^{\dagger}$ and $\mathcal{S}_{\mathrm{incoh}}^{\dagger} \mathcal{S}_{\mathrm{incoh}}$ possess the same eigenvalues that are given by 
\begin{equation}
\lambda_1=1+\sin(\alpha_1-\alpha_2), \ \lambda_2=1-\sin(\alpha_1-\alpha_2). 
\label{eigenvalues_simple}
\end{equation}
Therefore, our toy model shows that the deviation of the eigenvalues of the product $\mathcal{S}\mathcal{S}^{+}$ from unity is determined by the phase uncertainty after one scattering event, and hence it is directly related to the strength of decoherence. Moreover, those deviations are independent on the parameters $r$ and $t$ characterizing the coherent potential scattering in the chosen model. 

\section{Exact solution for the electron scattering probabilities averaged over magnetic impurities}
\label{exactSolution}

We study the effect of spin-flip scattering by magnetic impurities on the IQHE transition.  
 We adopt the model of point-like exchange interaction between spins of impurities and electron spins $H_{\mathrm{int}}= J \ {\bf I}\cdot{\bf s}$, where ${\bf I}$ and ${\bf s}$ denote the spins of impurities and of the electron respectively. Throughout the paper we assume spin-1/2 impurities.  In the absence of spin-flip scattering there are two Zeeman-split critical energies for each Landau level, where the QH delocalization transition takes place.  In Ref. \cite{kag-ch11}  it was found that the spin-flip scattering results in the appearance of a finite region of delocalized states around the critical QHE states. In the present paper we provide an analytical estimation of the width of the inter-plateaux transition based on the evaluation of coherence length due to scattering by magnetic impurities. 

In general, scattering of electrons by impurity spins induces many-electron Kondo correlations. In this paper however we consider regime, when the Kondo temperature is very low and Kondo correlations are suppressed. 
Scattering of electron by the saddle-point potential  in strong perpendicular magnetic field and in the presence of a magnetic impurity was studied in Ref. \cite{kag-ch11}. 

\begin{figure}
\includegraphics[width=8cm,height=11cm,angle=0]{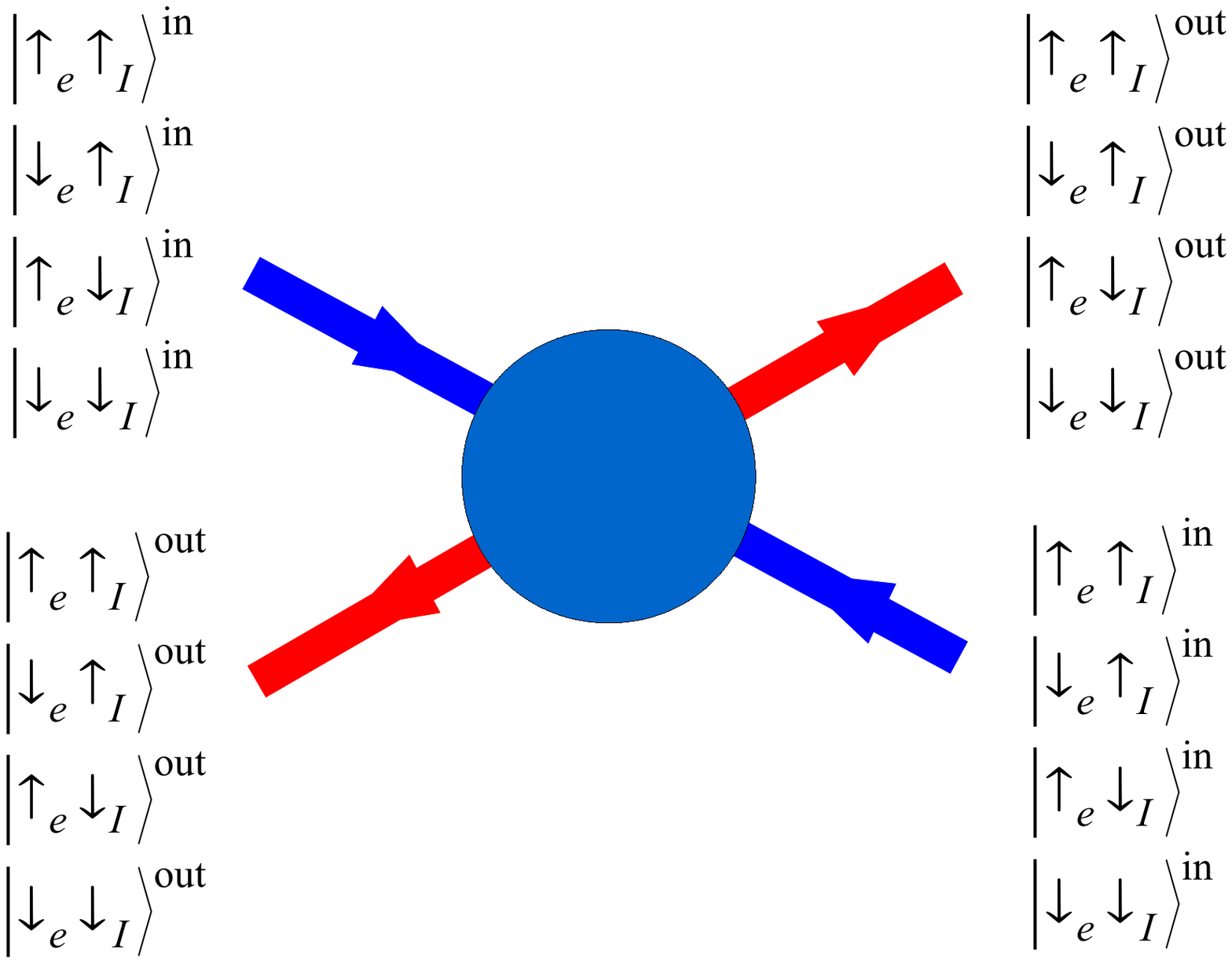}%
\vskip -4.cm
\caption{(Color online) Incoming and outgoing states at a single node.  Up and down arrows indicate $z$-components of the electron (subscript {\em e}) and impurity (subscript {\em I}) spins correspondingly. 
\label{fig-Node}}
\end{figure}

Following Ref. [\onlinecite{FH}], we introduce the dimensionless measure of energy $\epsilon=(E+\frac{1}{4}J)/E_1$, where $E_1$ is the energetic parameter characterizing the form of the saddle point potential. Furthermore, we denote the dimensionless  strength of exchange interaction as $\delta=J/E_1$.  That interaction results into two exchange-split energies $\epsilon_{1,2}=\epsilon\mp \delta/2$.  
Using the expressions for transmission and reflection coefficients, we construct the scattering matrix in the node relating the incoming and outgoing waves (see Fig. \ref{fig-Node}). 
\begin{equation}
S=\left(\begin{array}{cc} 
       \mathcal{R} & \mathcal{T} \\
      -\mathcal{T} & \mathcal{R}
      \end{array}
\right) = 
\left(
\begin{array}{rrrrrrrr}
r_1 & 0 & 0 & 0 & t_1 & 0 & 0 & 0 \\
0 & s_{22} & s_{23} & 0 & 0 & s_{26} & s_{27} & 0 \\
0 & s_{23} & s_{22} & 0 & 0  & s_{27} & s_{26} & 0 \\	
0 & 0 & 0 & r_1 & 0 & 0 & 0 & t_1 \\ 
-t_1 & 0 & 0 & 0 & r_1 & 0 & 0 & 0 \\
0 & -s_{26} & -s_{27} & 0 & 0 & s_{22} & s_{23} & 0 \\
0 & -s_{27} & -s_{26} & 0 & 0 & s_{23} & s_{22} & 0 \\
0 & 0 & 0 & -t_1 & 0 & 0 & 0 & r_1 
\end{array}
\right).  
\label{S-reduced}
\end{equation}
where the $4\times 4$ blocks $\mathcal{R}$ and $\mathcal{T}$ describe the reflection and transmission amplitudes respectively. 
Here the following notations were introduced 
\begin{eqnarray}
t_{1,2}=\frac{1}{\sqrt{1+e^{-\pi\epsilon_{1,2}}}}, & r_{1,2}=\sqrt{1-t_{1,2}^2}, & 
\label{t-and-r}\\
s_{22}=(r_1+r_2)/2, & s_{23}=(r_1-r_2)/2, \\
s_{26}=(t_1+t_2)/2, & s_{27}=(t_1-t_2)/2.
\end{eqnarray}
The absolute value squared of the particular scattering matrix element in Eq. (\ref{S-reduced}) gives the quantum scattering probability between the corresponding initial and final states of electron and impurity. 
Given the density matrix of the impurity spin, one can calculate the scattering probability for the {\em electron only}, averaged over the states of impurity. 
In what follows we assume a density matrix of magnetic impurity in the diagonal form 
\begin{equation}
\rho_I=\mathrm{diag}(w_{\uparrow}, w_{\downarrow}).  
\label{rhoI}
\end{equation}
The difference $w_{\uparrow} - w_{\downarrow}$ denotes the polarization degree of the magnetic impurity. 
After averaging over magnetic impurities, the resulting system looses the quantum coherence, and it can be described in terms of scattering probabilities. 
Using the density matrix Eq. (\ref{rhoI}), the averaged probability of the electron entering in the state with spin $\sigma$ to be reflected (transmitted) into the state with spin $\sigma'$ is given by 
\begin{equation}
R_{\sigma'\sigma}=\sum_{s, s'}  \rho_I^{s s} |\mathcal{R}_{\sigma' s', \sigma s}|^2,  \  
T_{\sigma'\sigma}=\sum_{s, s'}  \rho_I^{s s} |\mathcal{T}_{\sigma' s', \sigma s}|^2. 
\end{equation}
where $s, s'$ denote the initial and final spin states of the impurity. Note that the averaging takes place only over the initial spin state of impurity. Finally, the averaged probability matrix for the electron takes the form 
\begin{equation}
P=\left(\begin{array}{cc}
         R & T \\ 
	 T & R
        \end{array}
\right), 
\label{Prob-matrix}
\end{equation}
 where 
\begin{equation}
 R=  
\left(\begin{array}{cc}
       w_{\uparrow} r_1^2+ w_{\downarrow} s_{22}^2 & w_{\uparrow} s_{23}^2  \\
 w_{\downarrow} s_{23}^2 &  w_{\downarrow} r_1^2+ w_{\uparrow} s_{22}^2 
      \end{array}
\right), \ 
T=  
\left(\begin{array}{cc}
w_{\uparrow} t_1^2+ w_{\downarrow} s_{26}^2 &  w_{\uparrow} s_{27}^2 \\ 
w_{\downarrow} s_{27}^2 & w_{\downarrow} t_1^2+ w_{\uparrow} s_{26}^2  
      \end{array}
\right). 
\label{Prob-RT}
\end{equation}

\section{Introduction of fictitious scattering matrix}
\label{S_fictitious}

We now define the fictitious scattering matrix for quantum mechanical amplitude of the electron, which corresponds to the exact probability matrix obtained after the averaging over the states of magnetic impurity. To achieve that, we construct a scattering matrix with elements satisfying the following condition: the absolute square of each element must be equal to the corresponding probability of the matrix Eq. (\ref{Prob-matrix}). Furthermore, we choose the opposite signs of the elements in the two off-diagonal blocks, which ensures that the scattering matrix becomes unitary in the absence of spin-spin interaction, that is for $\delta=0$. Schematically, the scattering matrix acquires the form 
\begin{equation}
\mathcal{S}=\left(\begin{array}{cc}
         \sqrt{R} & \sqrt{T} \\ 
	 -\sqrt{T} & \sqrt{R}
        \end{array}
\right), 
\label{Sfict}
\end{equation}
where the square root is taken element-wise.  

\begin{figure}
\includegraphics[width=6cm,height=5cm,angle=0]{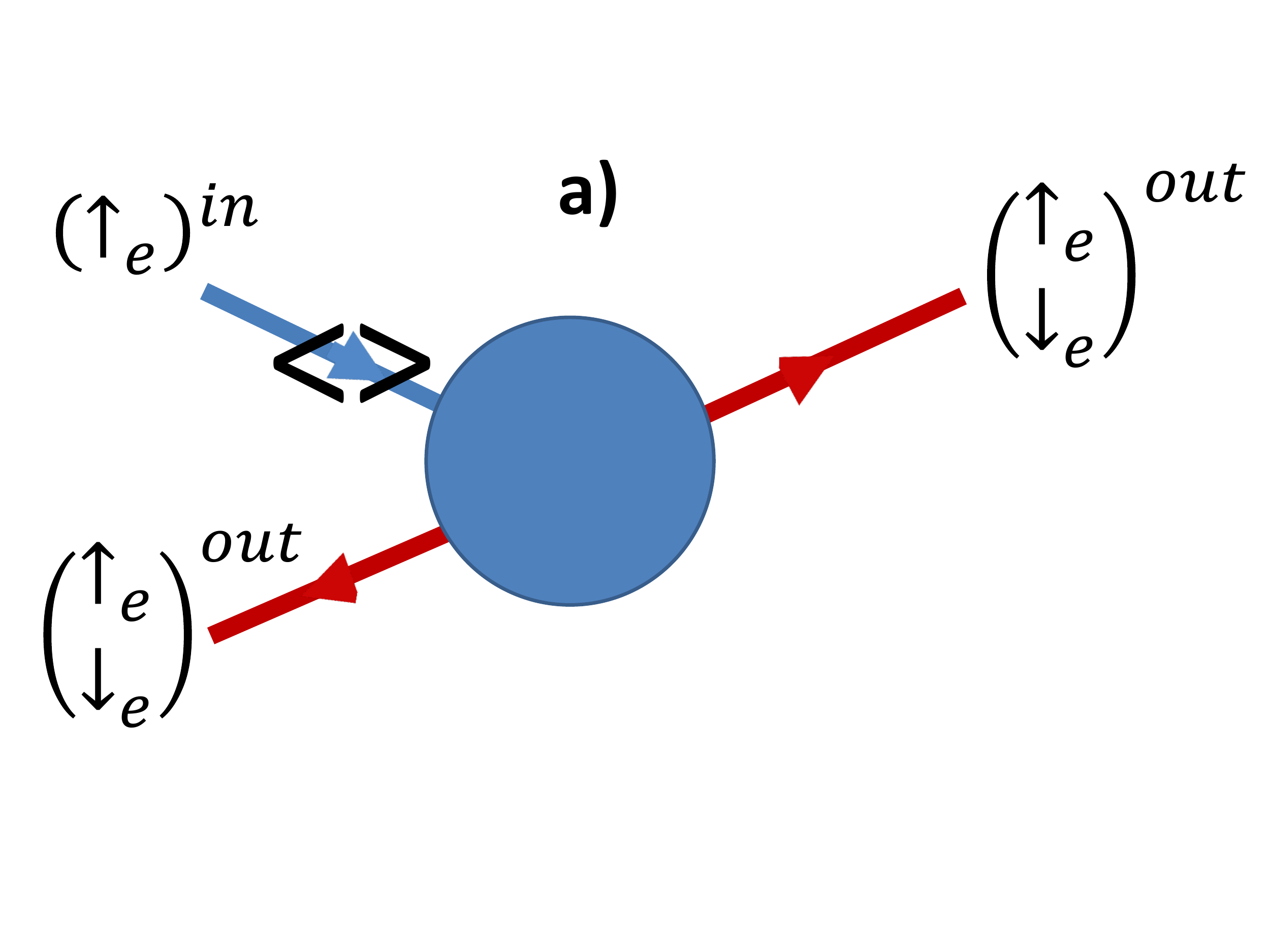}%
\hspace{2cm}
\includegraphics[width=6cm,height=5cm,angle=0]{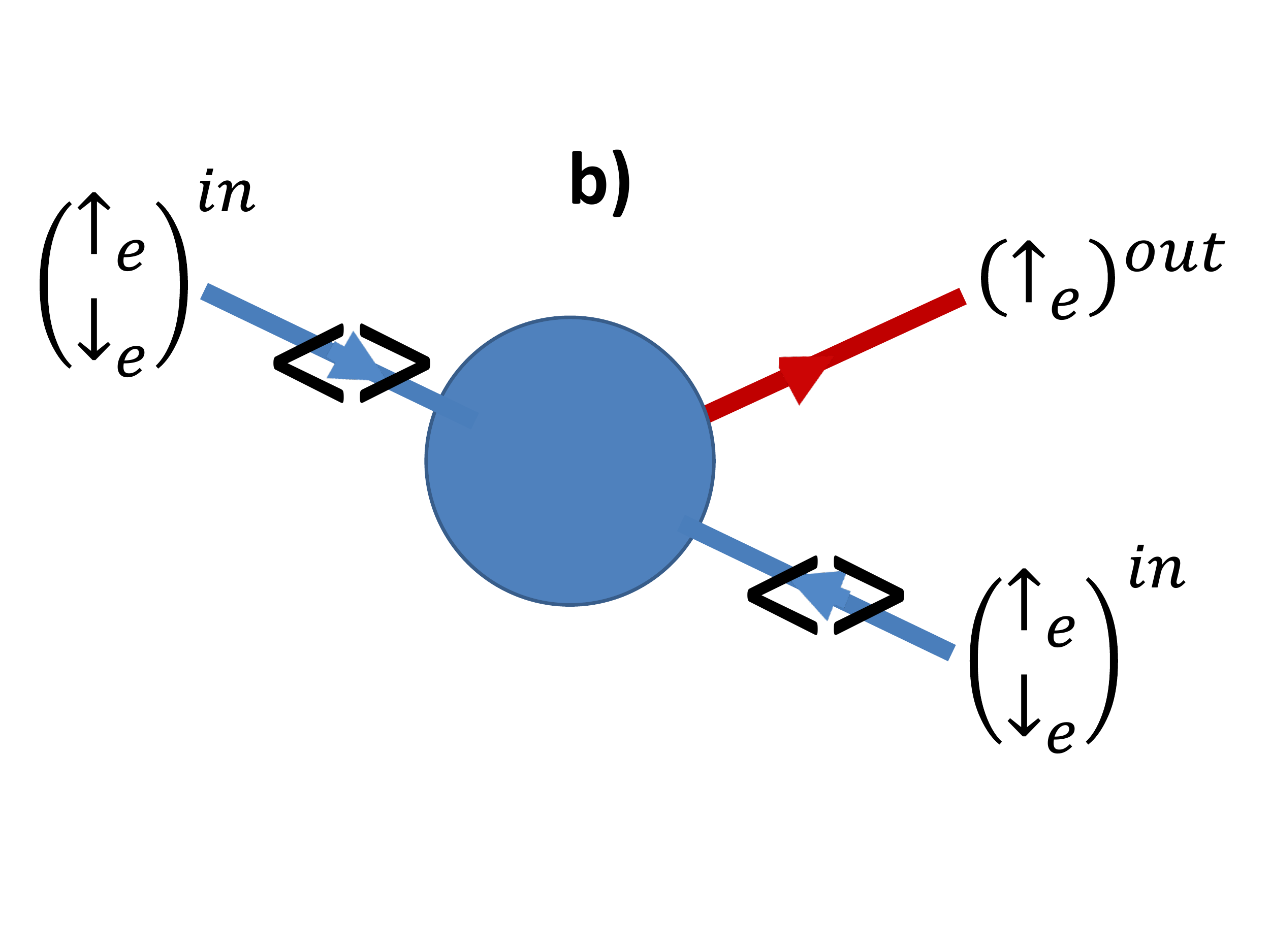}%
\vskip -1.cm
\caption{(Color online) a) Probability conservation as a sum of the elements squared of the first column in Eq. (\ref{Sfict}).  The incoming wave is scattered in the four (including spin)  outgoing channels.  b) Loss of the probability conservation by sum of the elements squared of the first row in Eq. (\ref{Sfict}).  The single outgoing wave in not the sum of the four incoming channels.   Angular brackets symbolize the averaging over the {\em initial} distribution of impurity spins. Because of the angular brackets, it is impossible to map the panel (b) onto panel (a), which is in contradistinction to the reversibility of quantum mechanics.  
\label{ProbOne}}
\end{figure}

Being nonunitary in general, the fictitious scattering matrix still possess some properties of a unitary matrix that follow from the conservation of probability. So, one can see from Eqs.  (\ref{Prob-matrix}), 
(\ref{Prob-RT}), (\ref{Sfict}), that the sum of the elements squared of each column in Eq. (\ref{Sfict}) equals 1, which describes the total probability for an electron entering the node to be scattered (see Fig. \ref{ProbOne}a). 
For example, the first column gives 
\begin{eqnarray}
\nonumber && 
       w_{\uparrow} r_1^2+ w_{\downarrow} s_{22}^2   +
 w_{\downarrow} s_{23}^2 +
w_{\uparrow} t_1^2+ w_{\downarrow} s_{26}^2 + 
w_{\downarrow} s_{27}^2 = w_{\uparrow}(r_1^2+t_1^2) + 
w_{\downarrow}(s_{22}^2+s_{23}^2+s_{26}^2+s_{27}^2)  \\
&& 
=w_{\uparrow}+w_{\downarrow}=1. 
\label{Prob-1}
\end{eqnarray}
The sum of the elements squared of each raw, which would correspond to the probability of a time-reversed scattering process,  differs from 1 (see Fig. \ref{ProbOne}b). This is due to the breaking of the time-reversal invariance introduced by averaging only over the initial states of the magnetic impurity.  For example, the sum of the elements of the first raw gives 
\begin{eqnarray}
\nonumber && 
       w_{\uparrow} r_1^2+ w_{\downarrow} s_{22}^2   +
 w_{\uparrow} s_{23}^2 +
w_{\uparrow} t_1^2+ w_{\downarrow} s_{26}^2 + 
w_{\uparrow} s_{27}^2 \\ 
\nonumber && 
= w_{\uparrow}(r_1^2+t_1^2) + 
w_{\downarrow}(s_{22}^2+s_{23}^2+s_{26}^2+s_{27}^2)+(w_{\uparrow}-w_{\downarrow})(s_{23}^2+s_{27}^2) \\
&& 
=1+(w_{\uparrow}-w_{\downarrow})(s_{23}^2+s_{27}^2). 
\label{Prob-not1}
\end{eqnarray} 
Note, that Eq. (\ref{Prob-not1}) gives unity in the case $w_{\uparrow}=w_{\downarrow}=1/2$, which corresponds to a completely unpolarized magnetic impurity. In that case, the time reversal symmetry appears to be restored. One can relate the restoration of the time reversal to the maximal possible entropy of the impurity spin, which, therefore, remains unchanged by the scattering and corresponds to a time reversible process in terms of thermodynamics. 

However, even in the case of the unpolarized impurity, the fictitious scattering matrix is not unitary because of decoherence introduced by averaging over the magnetic impurity. Formally, the different rows and columns of the matrix $\mathcal{S}$ are not orthogonal. This is the manifestation of violation of the orthogonality of two quantum states by phase decoherence (the toy model for that process is discussed in Sec. \ref{toyModel}).  

Now we apply the analysis of Section \ref{toyModel} to the fictitious scattering matrix Eq. (\ref{Sfict}). 
The nonunitary matrix $\mathcal{S}$ does not commute with its hermitian conjugated $\mathcal{S}^{\dagger}$. However, it is easy to show that the products $\mathcal{S}^{\dagger}\mathcal{S}$ and $\mathcal{S}\mathcal{S}^{\dagger}$ have the same eigenvalues. 
Calculating the eigenvalues of the product $\mathcal{S}^{\dagger}\mathcal{S}$ we obtain two doubly degenerate eigenvalues that can be cast into the form 
\begin{equation}
 \lambda_{1,2}=1\pm \sqrt{a^2+b^2}, 
\label{Eigenvalues_exact} 
\end{equation}
where $a=\left(\mathcal{S}^{\dagger}\mathcal{S}\right)_{12}$, and $b=\left(\mathcal{S}^{\dagger}\mathcal{S}\right)_{14}$. 
Note that the eigenvalues are symmetric with respect to unity. 
In the limit of weak spin-spin interaction, $\delta\ll 1$, the deviation of the eigenvalues from unity is given by 
\begin{equation}
c=\sqrt{a^2+b^2}\approx\frac{\pi\delta}{4} r_0t_0\left[\left(\sqrt{w_{\uparrow}}-\sqrt{w_{\downarrow}}\right)^2 
+\frac{\pi^2\delta^2}{16}r_0^2t_0^2\left(w_{\uparrow}^{3/2}+w_{\downarrow}^{3/2}\right)^2\right]^{1/2}, 
\label{c}
\end{equation} 
where $r_0$ and $t_0$ denote the reflection and transmission amplitudes Eq. (\ref{t-and-r}) calculated for $\delta=0$. 
According to the arguments given in Sec. \ref{toyModel}, the parameter $c$ serves as a measure of the decoherence introduced by the magnetic impurity. Moreover, comparing Eqs. (\ref{c}) and (\ref{eigenvalues_simple}), we conclude that $c$ measures the phase uncertainty acquired after a single incoherent scattering event. 
For the finite polarization of impurity the decoherence parameter $c$ grows linearly with $\delta$, 
\begin{equation}
c\approx \frac{\pi\delta}{4} r_0t_0 (\sqrt{w_{\uparrow}}-\sqrt{w_{\downarrow}}).  
\label{c_polarized}
\end{equation}
The dependence on $\delta$ becomes stronger with the degree of polarization of the impurity.  

\begin{figure}
\includegraphics[width=10cm,height=9cm,angle=0]{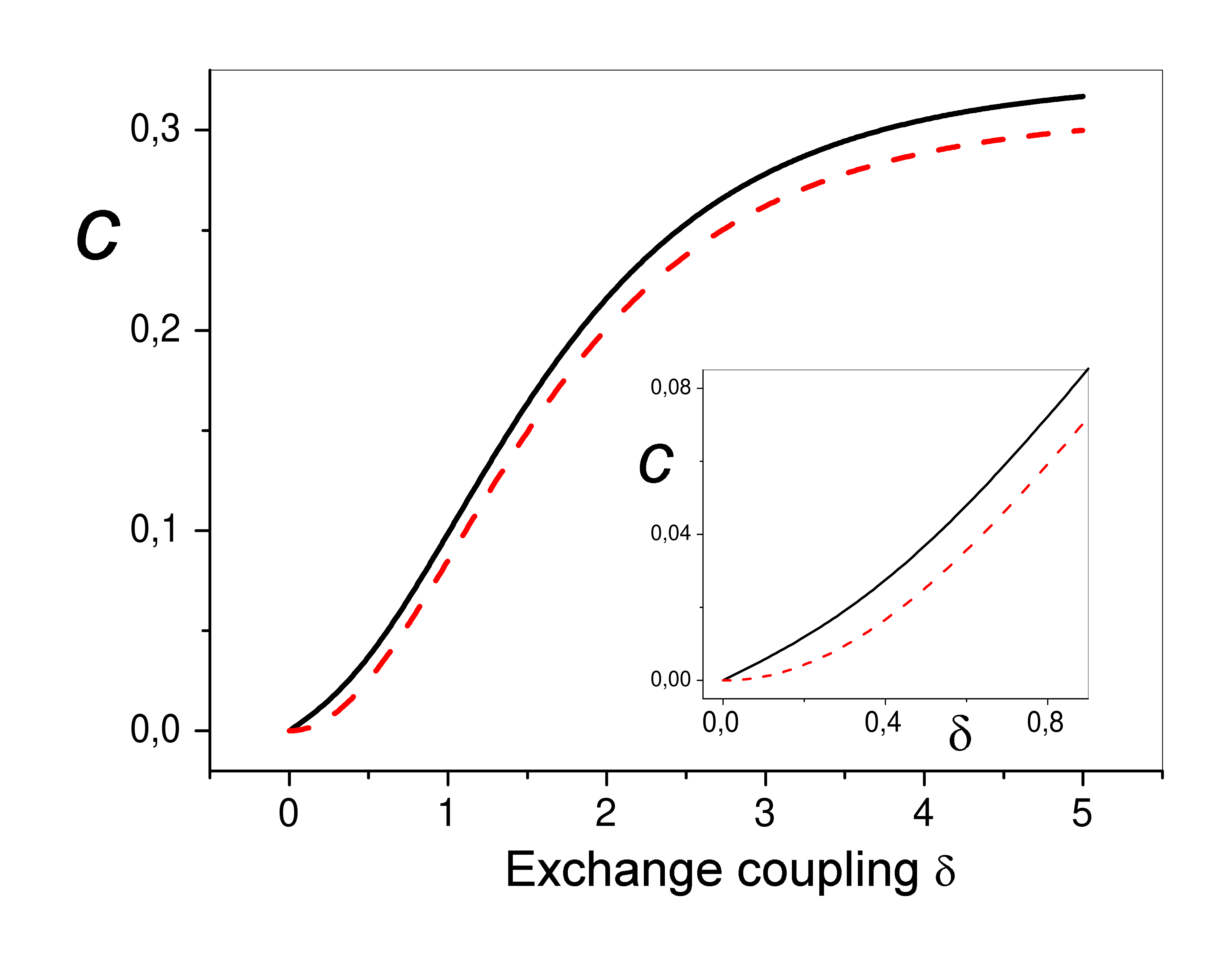}%
\vskip -0.5 cm
\caption{(Color online) Decoherence parameter $c$ as a function of the exchange coupling strength $\delta$. Energy $\epsilon=0$. Solid line: polarization $w_{\uparrow}-w_{\downarrow}=0.2$. Dashed line: polarization $w_{\uparrow}-w_{\downarrow}=0$. 
Inset shows the details of behavior $c(\delta)$ at small $\delta$. 
\label{c_delta}}
\end{figure}

In contrast, for the completely unpolarized impurity ($w_{\uparrow}=w_{\downarrow}=1/2$), the decoherence parameter $c$ grows with $\delta$ much slower, as $\delta^2$, 
\begin{equation}
c\approx \frac{\pi^2\delta^2r_0^2t_0^2}{16\sqrt{2}}. 
\label{c_unpolarized}
\end{equation}
This result is in accord with the restoration of the time reversal invariance of the fictitious scattering matrix Eq. (\ref{Sfict}) for the unpolarized impurity, which decreases the decoherence.  
Fig. \ref{c_delta} shows the dependence of the decoherence parameter $c$, as given Eq. (\ref{Eigenvalues_exact}),  on the exchange strength $\delta$ for the completely unpolarized ( $w_{\uparrow}-w_{\downarrow}=0$, dashed line), and for a weakly polarized ($w_{\uparrow}-w_{\downarrow}=0.2$, solid line) magnetic impurity.  The  dependence for small  $\delta\ll 1$ is shown in the inset in details. According to Eq. (\ref{c_unpolarized}), one  observes a purely quadratic dependence for the unpolarized impurity (dashed line). For a weak polarization,  a solid line exhibits a transition from the linear part in accord with Eq. (\ref{c_polarized}), to the nonlinear behavior at larger $\delta$, as described by Eq. (\ref{c}). 
Fig. \ref{c_delta} shows that the decoherence parameter $c$ saturates at large values of $\delta$. 

\begin{figure}
\includegraphics[width=10cm,height=9cm,angle=0]{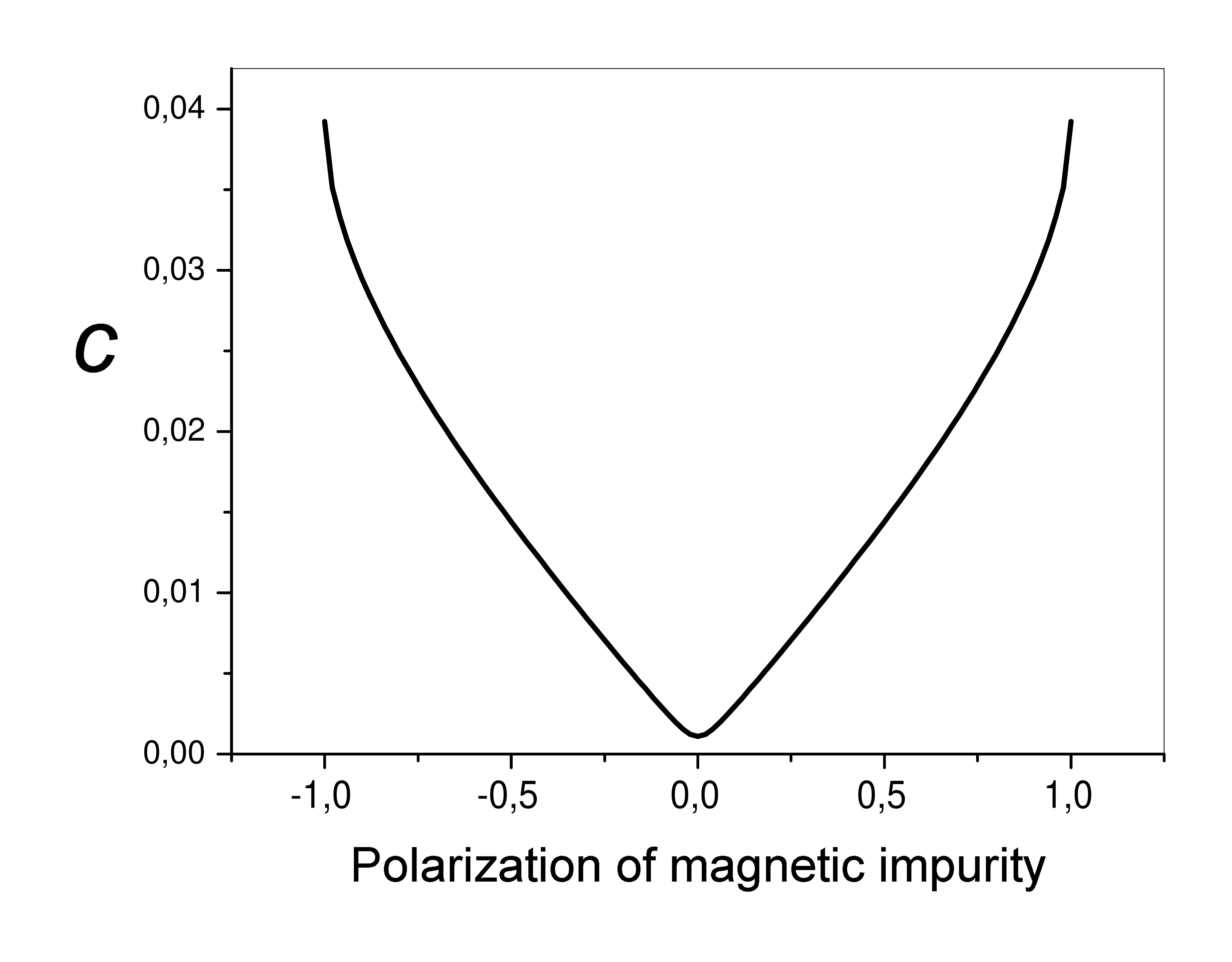}%
\vskip -0.5 cm
\caption{(Color online) Decoherence parameter $c$ as a function of the impurity polarization $w_{\uparrow}-w_{\downarrow}$.  Energy $\epsilon=0$, exchange coupling $\delta=0.1$.  
\label{c_w}}
\end{figure}

The dependence of decoherence parameter $c$ on the polarization of the magnetic impurity in shown in Fig. \ref{c_w}. 
According to previous discussion, the decoherence is minimal for the completely unpolarized impurity, and it grows monotonously with the impurity polarization. 

\section{Phase coherence length and broadening of inter-plateaux transition due to magnetic impurities} 
\label{plateaux}

We now apply the results of the previous section to the estimation of the phase coherence length due to scattering by magnetic impurities.  Further we evaluate the energetic width of the metallic region appearing at the inter-plateaux transition in the integer quantum Hall effect . 

The phase coherence length can be defined as a length of a path after which the phase uncertainty from multiple collisions becomes of the order of 1. Since a phase uncertainty in a single acte of scattering is a random quantity, the parameter $c$  evaluated in the previous section in Eq. (\ref{c})  should be understood as the dispersion of  the distribution of random scattering phases, 
\begin{equation}
 c=\sqrt{\left\langle \delta\phi^2\right\rangle} . 
\label{c_as_dispersion}
\end{equation}
The total phase uncertainty after multiple scattering events is evaluated as a sum of random phases, and it is given by 
\begin{equation}
\left\langle \delta \phi^2\right\rangle_N=N\left\langle \delta\phi^2\right\rangle=N c^2,  
\label{phi_N}
\end{equation}
where $N$ denotes the number of scattering events. Therefore, the number of scattering events needed to reach a complete decoherence is given by the relation $N c^2\sim 1$, hence $N\sim 1/c^2$. The corresponding phase coherence time can be estimated analogously to the calculation of the spin relaxation time by spin-orbit scattering due to Elliot-Yafet mechanism    \cite{Elliot-Yafet}
\begin{equation}
 \tau_{\mathrm{\phi}}\sim N \tau_{0}\sim \tau_{0}/c^2, 
\label{tauphi}
\end{equation}
where $\tau_{0}$ denotes the time between two consecutive scatteting events. The time $\tau_0$ is proportional to the distance between impurities. For a two-dimensional quantum Hall system $\tau_0\propto n_{\mathrm{imp}}^{-1/2}$, where $n_{\mathrm{imp}}$ denotes the concentration of magnetic impurities. 

Note that it follows from Eqs. (\ref{c_polarized}), (\ref{c_unpolarized}) that the inverse phase coherence time $1/\tau_{\phi}\propto c^2$ exhibits a crossover as a function of the exchange strength $\delta$   from the behavior $1/\tau_{\phi}\propto \delta^4$ for unpolarized magnetic impurities to $1/\tau_{\phi}\propto \delta^2$ if the magnetic polarization is finite. The crossover from $\delta^4$ behavior in the unpolarized system to the $\delta^2$ dependence for a finite spin polarization ($w_{\uparrow}\neq w_{\downarrow}$ is in accord with the findings of previous works Refs. \cite{Glazman-Kaminski,Altshuler}. 
The corresponding phase coherence length can be calcuated as a length of diffusion during the time $\tau_{\phi}$. 
\begin{equation}
 L_{\mathrm{\phi}}=\sqrt{D\tau_{\phi}}\sim \frac{1}{|c|n_{\mathrm{imp}}^{1/4}}. 
\label{Lphi}
\end{equation}

The region of the delocalized states in IQHE appears, when the phase coherence length for the electron becomes smaller than its localization length, which leads to the metallic behavior \cite{Zhao93,Polyakov93,kag95}.   
The phase coherence length of the electron corresponds to the length at which the phase uncertainty of its wave function becomes of the order of 1. 
  At the same time, close to the quantum Hall inter-plateaux transition, the localization length is known to scale with the deviation $\epsilon$ from the critical energy as $\xi\sim |\epsilon|^{-\nu}$ ($\nu\approx 2.6$)  \cite{nu}. Equating $L_{\mathrm{\phi}}$ and $\xi$, we obtain the estimation of the energetic width $\Delta$ of the metallic phase in the form 
\begin{equation}
 \Delta \sim \left(|c| n_{\mathrm{imp}}^{1/4}\right)^{\frac{1}{\nu}}.
\label{Delta_epsilon}  
\end{equation}
Therefore, substituting the results in Eqs. (\ref{c}), (\ref{c_polarized}), (\ref{c_unpolarized}) one obtains the dependence of the width of the metallic region on  both the strength of the spin-spin interaction and the polarization of magnetic impurities. 

\section{Summary and conclusions}
\label{sum}

In this paper we proposed a method of evaluation of the phase coherence length of an electron due to the scattering by magnetic impurities. The method is based on the introduction of the fictitious nonunitary scattering matrix that describes the electron motion averaged over the dynamics of magnetic impurities. The degree of nonunitarity is characterized by a single parameter $c$, which is the deviation of eigenvalues of the product $\mathcal{S}^{\dagger}\mathcal{S}$ from unity. The nonunitarity parameter is related to the phase uncertainty acquired in the single act of scattering, and it is inversely proportional to the phase coherence length. Our calculation revealed the change of the dependence of the nonunitarity parameter $c$ on the exchange coupling from the linear one at strong magnetic polarization to the quadratic one for the unpolarized magnetic impurities. 
 
With the help of the proposed method, we estimate the width of the metallic region at the IQHE inter-plateau transition and its dependence on the exchange coupling strength and the degree of polarization of magnetic impurities. We believe, that our method will be especially useful for other systems that allow the description in terms of scattering matrices and network models, such as topological insulators, graphene, quantum networks etc.  \cite{Furusaki,Levitov,Kimble}. 

\section{Acknowlengements}
We are grateful to I. Burmistrov for illuminating discussions. 
We  acknowledge financial support from the Collaborative Research Center (SFB) 668.  
This work was supported by the SCE under internal grant No. 5368911113.


\begin{thebibliography}{99}
\bibitem{Glazman-Kaminski} A. Kaminski and L. I. Glazman, Phys. Rev. Lett. {\bf 86}, 2400 (2001). 

\bibitem{Kumar2004} S. Kumar and P. Majumdar, Europhys. Lett., {\bf 65}, 75 (2004). 

\bibitem{Ramakrishnan} P. A. Lee, T. V. Ramakrishnan Rev. Mod. Phys. {\bf 57}, 287 (1985).  

\bibitem{Altshuler-Aronov} B. L. Altshuler and A. G. Aronov, pp. 1-153 in ``Electron-electron interaction in disoredered conductors'', ed. by A. L. Efros and Pollak, Elsevier Science 1985. 

\bibitem{kag-ch11} V. Kagalovsky and A. L. Chudnovskiy,   J. Mod. Phys. {\bf 2}, 970-976 (2011).  

\bibitem{LidarWhaley2003}D. A. Lidar, K. B. Whaley, in "Irreversible Quantum Dynamics", F. Benatti and R. Floreanini (Eds.), pp. 83-120,  Springer Lecture Notes in Physics {\bf 622}, Berlin, (2003)

\bibitem{FH} H. A. Fertig and B. I. Halperin, Phys. Rev. B {\bf 36}, 7969 (1987).

\bibitem{Elliot-Yafet} R. J. Elliot, Phys. Rev. {\bf 96}, 266 (1954); Y. Yafet, in {\em Solid State Physics}, edited by F. Seitz and
D. Turnbull (Academic, New York, 1963), Vol. 13; D. Huertas-Hernando, F. Guinea, and Arne Brataas, Phys. Rev. Lett. {\bf 103}, 146801 (2009).

\bibitem{Altshuler} G. G\"oppert, Y. M. Galperin, B. L. Altshuler, and H. Grabert, Phys. Rev. B {\bf 66}, 195328 (2002); 
G. G\"oppert and H. Grabert, Phys. Rev. B {\bf 64}, 033301
(2001). 

\bibitem{Zhao93} H. L. Zhao and S. Feng, Phys. Rev. Lett. {\bf 70}, 4134 (1993). 

\bibitem{Polyakov93} D. G. Polyakov and B. I. Shklovskii, Phys. Rev. Lett. {\bf 70}, 3796 (1993). 

\bibitem{kag95} V. Kagalovsky, B. Horovitz, and Y. Avishai, Europhys. Lett. {\bf 31}, 467 (1995). 

\bibitem{nu} H. Obuse, I. A. Gruzberg, and F. Evers
arXiv:1205.2763; K. Slevin, T. Ohtsuki, Phys. Rev. B {\bf 80}, 041304(R) (2009); M. Amado, A. V. Malyshev, A. Sedrakyan, and F. Domi­nguez-Adame, Phys. Rev. Lett. {\bf 107}, 066402 (2011). 

\bibitem{Furusaki} H. Obuse,  A. Furusaki, S. Ryu, and C. Mudry, Phys. Rev. B {\bf 76}, 075301 (2007). 

\bibitem{Levitov} D. A. Abanin, K. S. Novoselov, U. Zeitler, P. A. Lee, A. K. Geim, and L. S. Levitov, 
Phys. Rev. Lett. {\bf 98}, 196806 (2007). 

\bibitem{Kimble} H. J. Kimble, Nature {\bf 453},  1023, (2008). 




 
\end{thebibliography}
\end{document}